\documentclass[11pt]{article} 
\begin{document}
\input amssym.tex

\title{Collective classical motion on hyperbolic spacetimes of any dimensions}

\author{Ion I. Cot\u aescu\\        
{\small \em West University of Timisoara}\\{\small \em V. Parvan Ave. 4, RO-300223 Timisoara, Romania}}

\maketitle

\begin{abstract}
The geodesics equations on de Sitter and anti-de Sitter spacetimes of any dimensions, are the starting point for deriving the general form of the Boltzmann equation in terms of conserved quantities. The simple equation for the non-equilibrium Marle and Anderson-Witting models are derived and the distributions of the Boltzmann-Marle model on these manifolds are written down  first  in terms of conserved quantities and then as functions of canonical variables. 

Pacs: 04.20.Cv
\end{abstract}

Keywords: hyperbolic spacetimes; conserved quantities; geodesics; Boltzmann equation; Marle model; Anderson-Witting model; analytic solution.

\section{Introduction}
\label{ s1}

The  $(1+3)$-dimensional hyperbolic manifolds, the de Sitter (dS) and anti-de Sitter (AdS) spacetimes,  are interesting because of their isometries which generate a large collection of conserved quantities which may be used for studying  classical systems on these manifolds \cite{cac1,cac2,CdS1,CdS2,CdSgeo,CAdS1,CAdS2}.  

In our investigations we observed that on dS and AdS spacetimes there are {\em special} static charts \cite{P1,P2} with Cartesian coordinates in which the geodesic equations can be expressed simply, in terms of conserved quantities, without using explicitely initial conditions \cite{CdS2,CdSgeo,CAdS2}. These results can be  generalized easily to any dimensions by using the same type of static charts. Thus we  obtain the geodesic  equations on $(1+d)$-dimensional dS and AdS spacetimes in terms of conserved quantities with a plausible physical meaning, e. g. energy, angular momentum, etc.. 

With this starting point we focus on the collective classical motion on these manifolds  governed by the relativistic Boltzmann equation \cite{Cer} which is of actual interest in  relevant non-equilibrium models \cite{G1,G2,B1,B2}.  We  apply our previous results for writing down  the relativistic Boltzmann equation on the hyperbolic manifolds in terms of conserved quantities instead of the canonical coordinates. Thus we derive for the first time simple non-equilibrium equations depending only on time for  the Marle and Anderson-Witting models. Moreover, we succeed to solve analytically the distributions of  the Boltzmann-Marle models on the hyperbolic spacetimes of any dimensions. This is the principal result reported here which could be used in various applications.

We start in the second section presenting,   the $(1+d)$-dimensional dS and AdS spacetimes and their isometries, introducing the special static charts. Furthermore,  we derive the conserved quantities and the geodesic equations on these spacetimes. The third section is devoted to the dS and AdS mesoscopic systems governed by the Boltzmann equation.  Exploiting the previous results, we simplify this equation considering the distributions as functions on time and $2d$ conserved quantities instead of $2d$ canonical variables. In this manner we generalize our previous result \cite{AC} finding the form of the Maxwell-J\" uttner equilibrium distributions depending on the  dS and AdS local temperatures. The next section is devoted to the Marle and Anderson-Witting non-equilibrium models for which we write down simple equations and solve analytically  the Boltzmann-Marle equation giving the non-equilibrium  time-dependent distributions on the dS and AdS spacetime, first in terms of conserved quantities and then in canonical variables.    

\section{Particles on hyperbolic spectimes}
\label{s2}

The classical geodesic motion on dS and AdS spacetime can be studied exploiting the rich sets of conserved observables associated to their isometry groups. We shall show that with their help and by choosing {\em special} static charts  with suitable Cartesian coordinates \cite{P1,P2} we can write down the general form of the geodesic equations in arbitrary dimensions.

The $(1+d)$-dimensional dS and AdS spacetimes can be  defined as hyperboloids of radius $1/\omega$ embedded  in the same $[2+(d+1)]$-dimensional flat manifold $(M,\eta)$ of coordinates $z^A$, labeled by the indices $A,\,B,...= -1,0,1,2,..,d+1$, whose pseudo-Euclidean metric reads
\begin{equation}
\eta={\rm diag}(1,1, \underbrace{-1,-1,..-1}_{d+1})\,.
\end{equation}
The $(1+d)$-dimensional dS spacetime, denoted simply $dS(1+d)=M_+$, is defined as the intersection of the null cone of equation $z^2=\eta_{AB}z^A(x) z^B(x)=0$ with the subspace of fixed $z^{-1}=\frac{1}{\omega}$. Similarly we define the $(1+d)$-dimensional AdS spacetime, $AdS(1+d)=M_-$,  as the intersection of the null cone with the subspace given by $z^{d+1}=\frac{1}{\omega'}$. 

The embedding manifold $M$ carries the fundamental representation of the  group $G=SO(2,d+1)$ whose transformations ${\frak g}\in G$ change the Cartesian coordinates of $M$ as $z\to {\frak g}z$. According to our above definitions, it is obvious that the subgroup $G_+=SO(1,d+1)=I(M_+)$ is the isometry group of $M_+$ while $G_-=SO(2,d)=I(M_-)$ plays the same role in $M_-$. The group $G_0=G_+\bigcap  G_-=SO(1,d)$ is the gauge group of the metric 
\begin{equation}
\eta_0={\rm diag}(1,\underbrace{-1,-1,...-1}_{d})
\end{equation}
of the flat model $M_0$ of the spacetimes $M_+$ and $M_-$ whose isometry group, $I(M_0)=T(1+d)\circledS G_0$, generalizes the Poincar\' e group of the case $d=3$. Note that the hyperbolic spacetimes  have the coset structure $M_{\pm}\sim G_{\pm}/G_0$ \cite{Nach} we have exploited for building the dS and AdS relativity \cite{CAdS2,CdS1,CdS2}.   

For the group $G$ and its subgroups we use the canonical parametrization
\begin{equation}
{\frak g}(\xi)=\exp\left(-\frac{i}{2}\,\xi^{AB}{\frak S}_{AB}\right)\in SO(2,d+1) 
\end{equation}
with skew-symmetric parameters, $\xi^{AB}=-\xi^{BA}$,  and the covariant generators ${\frak S}_{AB}$ of the fundamental representation of the $so(2,d+1)$ algebra carried by $M$. In the Cartesian basis of $M$ these generators have the matrix elements, 
\begin{equation}
({\frak S}_{AB})^{C\,\cdot}_{\cdot\,D}=i\left(\delta^C_A\, \eta_{BD}
-\delta^C_B\, \eta_{AD}\right)\,.
\end{equation}
According to our previous interpretations, we say that the dS energy is  ${\frak H}_+= \omega{\frak S}_{0,d+1}$ while ${\frak H}_-= \omega{\frak S}_{-1,0}$ is the AdS one. In both these cases, the $so(d)$ generators,  ${\frak J}_{ij}={\frak S}_{ij}$, play the role of a generalized angular momentum while  ${\frak K}_i={\frak S}_{0,i}$ generalize the Lorentz boosts such that the set $({\frak J}_{ij},{\frak K}_i)$ forms a basis of the $so(1,d)$ Lie algebra of $G_0$.  A  Runge-Lenz-type vector of components ${\frak R}_i={\frak S}_{i,d+1}$ generate the $SO(d+1)/SO(d)$ transformations of $G_+$ while ${\frak N}_i ={\frak S}_{-1,i}$ generate supplemental boost transformations in $M_-$.

The local charts $\{x\}$  of coordinates $x^{\mu}$  (with Greek indices,  $\alpha,...\mu,\nu,...  =  0,1,2,...d$) can be introduced on the spacetimes $M_{\pm}$ giving the set of functions $z^A(x)$ which solve the above conditions. Here we consider only Cartesian space coordinates which satisfy the condition  $z^i\propto x^i$  such that at least the $SO(d)$ symmetry becomes global, any quantity carrying space indices  $i,j,k,...=1,2,...d$ transforming as  $SO(d)$  vectors and tensors.  For this reason we shall use some Euclidean notations as the scalar product $(a\cdot b)=a_ib_i$ of the vectors $a,b\in{\Bbb R}^d$ and the squared Euclidean norm,   $a^2=(a\cdot a)$. 

In what follows we use only  the {\em special} static chars, $\{x_+\}=\{t_+,{x}\}$ on $M_+$ and  $\{x_-\}=\{t_-,{x}\}$ on $M_-$, having the same space coordinates for pointing out the  symmetry of the spacetimes $M_+$ and $M_-$ with the same radius $\frac{1}{\omega}$ ($\omega'=\omega$). The embedding equations defining the special static charts of these manifolds are
\begin{eqnarray}
M_+ &=& dS(1+d): \quad M_-=AdS(1+d):\nonumber\\
%&&\nonumber\\
z_{+}^{-1}&=&\frac{1}{\omega}\hspace*{19mm}z_{-}^{-1}=\frac{\cos\omega t_-}{\omega\chi_-}\\
z_{+}^{0}&=&\frac{\sinh\omega t_+}{\omega\chi_+} \hspace*{10mm}z_{-}^{0}=\frac{\sin\omega t_-}{\omega\chi_-}\\
z_{+}^i&=&\frac{x^i}{\chi_+}\hspace*{18.5mm}z_{-}^i=\frac{x^i}{\chi_-}\\
z_{+}^{d+1}&=&\frac{\cosh\omega t_+}{\omega\chi_+}\hspace*{7mm}z_{-}^{d+1}=\frac{1}{\omega}
\end{eqnarray}
where $\chi_\pm =\sqrt{1\pm\omega^2 x^2}$. The corresponding line elements can be calculated according to the general rule as 
\begin{eqnarray}
ds_{\pm}^2&=&\eta_{AB}dz^A_{\pm}dz^B_{\pm}=g^{\pm}_{\mu\nu}dx_{\pm}^{\mu}dx_{\pm}^{\nu}\nonumber\\
&=&\frac{1}{\chi_{\pm}^2}\left[dt_{\pm}^2-\left(\delta_{ij}\mp\frac{\omega^2}{\chi_{\pm}^2}x^i x^j\right)dx^i dx^j\right]\,.
\end{eqnarray} 
Note that the special static charts introduced here are not quite popular since in many applications one prefers the standard static charts $\{t_{\pm}, x_{s\pm}\}$ with static Cartesian coordinates $x^i_{s\pm}$ related to our coordinates $x^i$ as in the Appendix A. 

The transformation ${\frak g}_{\pm}\in G_{\pm}$ generates the isometry  $x_{\pm}\to x_{\pm}'=\phi_{{\frak g}_{\pm}}(x_{\pm})$ derived from system of equations  $z[\phi_{\frak g}(x)]={\frak g}z(x)$ that have to be solved in a given chart of $M_+$ or $M_-$. These isometries give rise to the principal conserved quantities associated to the $so(2,d+1)$ generators. The conserved quantities along the geodesics of $M_{\pm}$ are given by the Killing vectors associated to the $G_{\pm}$  isometries. In the charts  $\{x\}_{\pm}$ of $M_+$ or $M_-$ the components of the Killing vectors are defined (up to a multiplicative constant)  as \cite{ES},
\begin{equation}
k^{\pm}_{(AB)\,\mu}=z_{\pm\,A}\partial_{\mu} z_{\pm\, B} -z_{\pm\, B}\partial_{\mu} z_{\pm\,A}\,,\quad  \,.
\end{equation}
where $z_{\pm\, A}=\eta_{AC}z_{\pm}^C$. Then the conserved quantities along a timelike geodesic of a particle of mass $m$ have the form $k^{\pm}_{(AB)\,\mu}p_{\pm}^{\mu}$ where 
\begin{equation}\label{cmom}
p_{\pm}^{\mu}=m\frac{dx_{\pm}^{\mu}}{ds_{\pm}}\,, 
\end{equation}
are the components of the covariant momenta which satisfy $p_{\pm}^2=g_{\mu\nu}^{\pm}p_{\pm}^{\mu}p_{\pm}^{\nu}=m^2$
in both the charts under consideration. 

With these preparations, we may express the equations of the timelike geodesics exclusively in terms of $2d$ conserved quantities which replace completely the $2d$ initial conditions  determining usually a geodesic in a $(1+d)$-dimensional manifold. 

We focus first on the dS spacetime calculating the components of the Killing vectors and defining with their help the conserved quantities,
\begin{eqnarray}
E^+&=&\omega k^+_{(0,d+1)\, \mu}p_+^{\mu}=\frac{p_+^0}{\chi_+^2}\,,\label{Eplus}\\
L_{ij}^+&=& k^+_{(i,j)\, \mu}p_+^{\mu}=\frac{x_+^i p_+^j-x_+^j p_+^i}{\chi_+^2}\\
K_i^+&=&k^+_{(0,i)\, \mu}p_+^{\mu}\nonumber\\
&=&\frac{x_+^i p_+^0}{\chi_+^2}\cosh\omega t_+ -\frac{ p_+^i}{\omega\chi_+^2}\sinh\omega t_+\label{Kplus}\\
R_i^+&=&k^+_{(i,d+1)\, \mu}p_+^{\mu}\nonumber\\
&=&\frac{x_+^i p_+^0}{\chi_+^2}\sinh\omega t_+ -\frac{ p_+^i}{\omega\chi_+^2}\cosh\omega t_+\label{Rplus}
\end{eqnarray}  
where $E^+$ is the energy and $L^+$ is the generalized angular momentum corresponding to the $SO(d)$ generators. The vector $K^+$ is associated to the generalized Lorentz boost of the group $G_0$ while the vector $R^+$ is specific for the dS isometries being associated to the space rotations involving the coordinate $z^{d+1}$.  These quantities are not independent since the generalized angular momentum can be written as 
\begin{equation}\label{Lplus}
L^+_{ij}=\frac{\omega }{E^+}\left(R^+_i K^+_j-R^+_j K^+_i\right)\,,
\end{equation} 
and we have the identity corresponding to the first Casimir invariant of the $so(1,d+1)$ algebra, 
\begin{equation}\label{Eidplus}
(E^+)^2-\omega^2\left[ (L^+)^2+(R^+)^2-(K^+)^2\right]=m^2\,,
\end{equation}
depending on the Euclidean squared norms $(L^+)^2=L^+_{ij} L^+_{ij}$, $(R^+)^2=R^+_i R^+_i$, etc.. Hereby we conclude that there are only $2d$ independent conserved quantities, $(K^+_i, R^+_i)$, which form a convenient algebraic basis generating freely all the other conserved quantities.

Now we can exploit the above results for expressing the geodesic equations on $M_+$ in terms of conserved quantities instead of using $2d$ arbitrary initial conditions.  Indeed, from Eqs. (\ref{Eplus}), (\ref{Kplus}) and (\ref{Rplus}) we obtain the geodesic equation in a closed form as
\begin{equation}\label{geoplus}
x^i (t_+)=\frac{1}{E^+}\left(K_i^+ \cosh \omega t_+ -R_i^+ \sinh \omega t_+\right)
\end{equation}
which represents a hyperbola in the plane $(K^+, R^+)$ whose asymptotes are oriented along the vectors 
\begin{equation}
\frac{1}{2\omega E^+}\left(K^+-R^+\right)\,, \quad \frac{1}{2\omega E^+}\left(K^++R^+\right)\,.
\end{equation} 
We recover thus our previous result obtained recently for $d=3$ \cite{CdSgeo}. In addition, we obtain the momentum components,
\begin{equation}
p^i_{+} (t_+)=\omega\chi_+^2 \left(K_i^+ \sinh \omega t_+ -R_i^+ \cosh \omega t_+\right)
\end{equation}
that can be used in applications.

In a similar manner we calculate the components of the Killing vectors in the chart $\{x_-\}$ of the SdS spacetime,  defining the conserved quantities,
\begin{eqnarray}
E^-&=&\omega k^-_{(-1,0)\, \mu}p_-^{\mu}=\frac{p_-^0}{\chi_-^2}\,,\label{Emin}\\
L_{ij}^-&=& k^-_{(i,j)\, \mu}p_-^{\mu}=\frac{x_-^i p_-^j-x_-^j p_-^i}{\chi_-^2}\\
K_i^-&=&k^-_{(0,i)\, \mu}p_-^{\mu}\nonumber\\
&=&\frac{x_-^i p_-^0}{\chi_-^2}\cos\omega t_- -\frac{ p_-^i}{\omega\chi_-^2}\sin\omega t_-\,,\label{Kmin}\\
N_i^-&=&k^-_{(-1,i)\, \mu}p_-^{\mu}\,,\nonumber\\
&=& -\frac{x_-^i p_-^0}{\chi_-^2}\sin\omega t_- -\frac{p_-^i}{\omega\chi_-^2}\cos\omega t_-\,.\label{Nmin}
\end{eqnarray}  
The quantities  $L^-$ and $K^-$ correspond to the $so(1,d)$ basis generators while $E^-$ is the AdS energy. The vector $N^-$ is  associated to the boosts involving the second time coordinate $z^{-1}$ which generate only AdS isometries. As in the previous case we  
find that
\begin{equation}
L^-_{ij}=\frac{\omega }{E^-}\left(N^-_i K^-_j-N^-_j K^-_i\right)\,,
\end{equation} 
while the identity corresponding to the first Casimir invariant of the $so(2,d)$ algebra reads
\begin{equation}
(E^-)^2+\omega^2\left[ (L^-)^2-(N^-)^2-(K^-)^2\right]=m^2\,.
\end{equation}
Therefore, the $2d$ independent conserved quantities $(K^-_i, N^-_i)$ represent a natural basis. Furthermore, according to Eqs. (\ref{Emin}), (\ref{Kmin}) and (\ref{Nmin}), we obtain the geodesic equation in a closed form,
\begin{equation}\label{geomin}
x^i (t_-)=\frac{1}{E^-}\left(K_i^- \cos \omega t_- -N_i^- \sin \omega t_-\right)\,,
\end{equation}
which represents an ellipse in the plane $(K^-, N^-)$ just as in the case of $d=3$ \cite{CAdS1,CAdS2}. Moreover,  we can derive the momentum components as
\begin{equation}\label{geomin1}
p^i (t_-)=-\omega \chi_-^2\left(K_i^- \sin \omega t_- +N_i^- \cos \omega t_-\right)\,.
\end{equation}

Finally we note that here we respected up to signs our previous definitions given in the cases of $d=3$ for the dS  \cite{CdS1,CdS2} and AdS  \cite{CAdS1,CAdS2} manifolds.  Therefore, we find similar flat limits (for $\omega \to 0$) that read 
\begin{eqnarray}
&&\lim_{\omega \to 0}\,E^{\pm}=E=p^0\,,\\
&&\lim_{\omega \to 0}\,L_{ij}^{\pm}=L_{ij}=x^ip^j-x^jp^i \,,\\
&&\lim_{\omega \to 0}\,K^{\pm}_i=K_i=x^i p^0-t p^i\,,\\
&&\lim_{\omega \to 0}\,\omega R^+_i=\lim_{\omega \to 0}\omega N^-_i=-p^i
\end{eqnarray} 
where $(p^{\mu}, L_{ij},K_i)$ are the conserved quantities on $M_0$ corresponding to the basis generators  of its isometry group $I(M_0)$. 

\section{Boltzmann equations on $M_{\pm}$}
\label{s3}

The above results allow us to continue our previous study \cite{AC} of the mesoscopic systems on $M_{\pm}$ considering distributions which depend only on time and $2d$ conserved quantities and, consequently,  satisfy simplified Boltzmann equations.

A mesoscopic system  constituted by identical particles of mass $m$ on a $(1+d)$-dimensional courved bakground  is successfully described  by the (general) relativistic Boltzmann equation \cite{Cer},
\begin{equation}
\frac{\partial f_B (x,p)}{\partial x^{\mu}}p^{\mu}-\Gamma_{\alpha\beta}^{i}p^{\alpha}p^{\beta}\frac{\partial f_B (x,p)}{\partial p^{i}}=J[f_B]\,,
\end{equation}
giving the scalar distribution $f_B(x,p)$ which depends on the local coordinates $x^{\mu}$  and momentum components  $p^{\mu}$ along geodesics. These satisfy the geodesic equation and the normalization condition $g_{\mu\nu}p^{\mu}p^{\nu}=m^2$ such that we remain only with $d$ independent momentum variables, say $p^i$, such that the function $f_B$ depends on $2d+1$ variables, i. e the time $t$ and $2d$ canonical variables, $(x^i, p^i)$.    

These variables can be changed at any time if we have  $n$  vectors fields $K^a,\, a,b,...=1,2,...n$  defining the new quantities $k^a=K^a_{\mu}(x)p^{\mu}$ that can play the role of new variables. Then we can substitute $n$ canonical variables with new ones by  solving the above system of $n$ equations for expressing $n$ canonical variables in terms of the new variables $k_a$ and the remaining $2d-n$ canonical variables. If $n>d$ we can replace all the momentum components by the functions $p^{\mu}(t,x,k)=p^{\mu}(t,x^1,x^2,..x^s,k^1,k^2,...k^n)$ where $s=2d-n$. In this manner we obtain a new function $f(t,x,k)$  depending on the new variables which satisfies now
\begin{equation}
\frac{df}{ds}=\frac{\partial f}{\partial x^{\mu}}u^{\mu}+\frac{\partial f}{\partial k^{a}}\frac{d k^{a}}{ds}\,.
\end{equation}
On the other hand, we observe that
\begin{equation}
\frac{dk^a}{ds}=K_{\mu}^a \frac{du^{\mu}}{ds}+\frac{\partial K^a_{\mu}}{\partial x^{\nu}}u^{\nu} u^{\mu}=K^a_{\mu;\nu}u^{\mu} u^{\nu}\,,
\end{equation}
since $u$ satisfy the geodesic equation. Thus we obtain the new general equation
\begin{equation}
\frac{\partial f}{\partial x^{\mu}}p^{\mu}+\frac{\partial f}{\partial k^{a}}K^a_{\mu;\nu}p^{\mu} p^{\nu}=J[f]\,,
\end{equation}
depending on the variables $(t,x,k)$. This procedure is useful  when $K^a$ are Killing vector fields since then the Killing equations $K^a_{\mu;\nu}p^{\mu}p^{\nu}=0$ drop out all the terms containing derivatives $\partial_{k_a}f$,  remaining with a simpler equation depending only on time and $s$ space coordinates. 

Important applications can be worked out on the hyperbolic spactimes which offer us just $n=2d$ independent conserved quantities allowing us to write on $M_{\pm}$ simple Boltzmann equations depending only on time,      
\begin{equation}\label{Bol}
\frac{\partial f_{\pm} (t_{\pm},k_{\pm})}{\partial t_{\pm}}p_{\pm}^{0}(t_{\pm},k_{\pm})=J[f_{\pm}]\,,
\end{equation}
where $k_+=(K^+_i,R_i^+)$ on $M_+$ and $k_-=(K^-_i,N^-_i)$ on $M_-$. The components
\begin{equation}\label{p0pm}
p^0_{\pm}=E^{\pm}\chi_{\pm}^2=E^{\pm}\left[1\pm\omega^2 x(t_{\pm})^2\right]
\end{equation}
have to be calculated by using the geodesic equations  (\ref{geoplus}) and (\ref{geomin}).

The equation (\ref{Bol}) can be solved now analytically  if the collision term $J$  is approximated according to the Marle or Anderson-Witting models. In both these cases the distributions  on $M_{\pm}$ have the general form
\begin{equation}\label{fff}
f_{\pm}=f_{\pm}^{(eq)}(k_{\pm}) +\delta f_{\pm}(t_{\pm},k_{\pm})\,,
\end{equation}
where $f^{(eq)}_{\pm}$ are the Maxwell-J\" uttner equilibrium distributions on $M_{\pm}$ while the corrections $\delta f_{\pm}$ giving the local transport effects have to be calculated from Eq. (\ref{Bol}) by using the above mentioned approximations.

Let us first analyze the  Maxwell-J\" uttner distributions on $M_{\pm}$ which must have the general form \cite{AC}
\begin{equation}
f_{\pm}^{(eq)}(k_{\pm})=\frac{Z}{(2\pi)^{d}}\exp \left[-\beta_E^{\pm}(x) p^{\mu}_{\pm} U^{\pm}_{\mu}\right]
\end{equation}
where $U$ is the macroscopic field of velocities of the macroscopic system. Since this distribution is independent on time, a rapid inspection shows that there is only one obvious choice, namely $U^{\pm}_i=0$ and, consequently, $U^{\pm}_0=(\chi_{\pm})^{-1}$.  Then, bearing in mind that  $p^0_{\pm}$ are given by Eq. (\ref{p0pm}), we obtain 
\begin{equation}\label{pU}
p^{\mu}_{\pm} U^{\pm}_{\mu}=p^{0}_{\pm} U^{\pm}_{0}=E^{\pm}\chi_{\pm}\,,
\end{equation}
such that the local temperature reads
\begin{equation}\label{temp}
\beta_E^{\pm}(x)=\frac{\beta_0}{\chi_{\pm}}=\frac{\beta_0}{\sqrt{1\pm \omega^2 x^2}}\,,
\end{equation}
where $\beta_0$ is the temperature in origin. Therefore, the final form of the 
Maxwell-J\" uttner distribution reads
\begin{eqnarray}
f_{\pm}^{(eq)}(k_{\pm})&=&\frac{Z_{\pm}}{(2\pi)^{d}}\exp (-\beta_0 E^{\pm}) \nonumber\\
&=&\frac{Z_{\pm}}{(2\pi)^{d}}\exp\left(-\frac{\beta_0 p^0_{\pm}}{1\pm \omega^2 x^2}\right)\,.\label{MJ}
\end{eqnarray}
This result is not surprising since it is natural to find that the mesoscopic systems which are  in equilibrium in  static charts must stay at rest as we have show recently in Ref. \cite{AC}.   

\section{Marle and Anderson-Witting models}
\label{s32}

The simplest approximation of the collision term of Eq. (\ref{Bol}) is given by the Marle model which defines 
\begin{equation}\label{Mar}
J_M[f_{\pm}]=-\frac{m}{\tau}\left(f_{\pm}-f_{\pm}^{(eq)}\right)\,,
\end{equation} 
where the relaxation time $\tau$ is a new parameter. Therefore, we may find the functions $\delta f_{\pm}^W$ by solving the equations 
\begin{equation}\label{Bol1}
E^{\pm}\frac{\partial\, \delta f_{\pm}^M (t_{\pm},k_{\pm})}{\partial t_{\pm}}\,\chi_{\pm}^{2}=- \frac{m}{\tau}\,\delta f_{\pm}^M (t_{\pm},k_{\pm})\,,
\end{equation}  
resulted from Eqs. (\ref{Bol}), (\ref{p0pm}) and  (\ref{Mar}). 

Another useful approximation of the collision term is given by the Anderson-Witting model where
\begin{equation}
J_{AW}[f_{\pm}]=-\frac{p^{\mu}_{\pm}U_{\mu}^{\pm}}{\tau}\left(f_{\pm}-f_{\pm}^{(eq)}\right)\,.
\end{equation}
Then, by taking into account that the quantity  $p^{\mu}_{\pm}U_{\mu}^{\pm}$ was already evaluated in Eq. (\ref{pU}), we obtain the equations
\begin{equation}\label{Bol2}
\frac{\partial\, \delta f_{\pm}^{AW} (t_{\pm},k_{\pm})}{\partial t_{\pm}}\,\chi_{\pm}=- \frac{1}{\tau}\,\delta f_{\pm}^{AW} (t_{\pm},k_{\pm})\,,
\end{equation}
satisfied by the distributions $\delta f_{\pm}^{AW}$ of this model. 

Obviously, in the case of the hyperbolic spacetimes, the Marle and Anderson-Witting models give different distributions depending on time.  In what follows we discuss the solution of both these models in the dS and AdS spacetimes separately such that the notation $\pm$ is no longer needed.

{\em Distributions in dS spacetimes.} Let us consider first the dS spacetime $M_+$ where we denote now by $\{t,x\}$ the  special static chart. Then by using algebraic codes on computer and Eqs. (\ref{Bol1}),      (\ref{Eidplus}) and (\ref{Lplus}) we find the distribution of the Marle model
\begin{eqnarray}
&&\delta f^M(t,K,R)=h(K,R)\nonumber\\
&&~~~~\times\exp\left[-\frac{1}{\omega\tau}{\rm arctanh}\left(A\tanh\omega t+B\right) \right]\,,
\end{eqnarray}
where
\begin{equation}
A=\frac{E^2-\omega^2 R^2}{mE}\,, \quad B=\frac{\omega^2 (K\cdot R)}{mE}\,.
\end{equation}
The quantity $h(K,R)$ which plays the role of an integration constant has to be specified according to the needs of the concrete physical model. When $\delta f$ is a small correction we can take  $h(K,R)=\kappa f^{(eq)}_+$ where the equilibrium distribution is given by Eq. (\ref{MJ}) while $\kappa$ is a new parameter. Thus we obtain the distribution (\ref{fff}) of the Marle model depending on $(t,E,K_i,R_i)$. The last step is to turn back to the canonical variables $(x^i,p^i)$  by substituting the conserved quantities according to Eqs. (\ref{Eplus}), (\ref{Kplus}) and (\ref{Rplus}). We obtain thus the definitive form
\begin{eqnarray}
&&\delta f^M(t,x,p)=h(K,R)\nonumber\\
&&\times\exp\left[-\frac{1}{\omega\tau}{\rm arctanh}\left(\frac{p^0}{m}\tanh\omega t-\frac{\omega}{m}\frac{(x\cdot p)}{1+\omega^2 x^2}\right) \right]\,,
\end{eqnarray} 
where $p$ is the covariant momentum (\ref{cmom}).

Unfortunately,  the Anderson-Witting equation  (\ref{Bol2})  cannot be integrated in the general case on $M_+$ by using only algebraic codes. This means that we must study more carefully this equation  resorting to refined mathematical methods, the last chance being the  use of the numerical ones. 

{\em Distributions in AdS spacetimes}
On the AdS spacetime $M_-$, in the special static chart denoted by $\{t,x\}$,  the  Eq.  (\ref{Bol1}) can be solved obtaining the Marle distribution  that reads
\begin{eqnarray}
&&\delta f^M(t,K,N)=h(K,N)\nonumber\\
&&~~~~~~\times \exp\left[-\frac{1}{\omega\tau}{\rm arctan}\left( A'\tan\omega t+B' \right) \right]\,,
\end{eqnarray}
where
\begin{equation}
A'=\frac{E^2-\omega^2 N^2}{mE}\,, \quad B'=\frac{\omega^2(K\cdot N) }{mE}\,,
\end{equation}
while the integration constant can be taken as $h(K,N)=\kappa f^{(eq)}_-$ or in accordance to other physical needs. Thus the Marle distribution on $M_-$ is completely determined by Eq. (\ref{fff}) and  we get back to the canonical variables  finding the final result,
\begin{eqnarray}
&&\delta f^M(t,x,p)=h(K,N)\nonumber\\
&&\times\exp\left[-\frac{1}{\omega\tau}{\rm arctan}\left(\frac{p^0}{m}\tan\omega t-\frac{\omega}{m}\frac{(x\cdot p)}{1-\omega^2 x^2}\right) \right]\,,
\end{eqnarray} 
resulted from  Eqs. (\ref{Emin}), (\ref{Kmin}) and (\ref{Nmin}).

As in the previous case there are major difficulties in integrating the Eq. (\ref{Bol2}) of the 
Anderson-Witting model which requires a careful study in each particular case separately.

\section{Conclusion}

This paper is a technical piece of work whose principal purpose was to find new general  results concerning the classical motion on hyperbolic spactimes of any dimensions. The generalization of the geodesic equations expressed in terms of conserved quantities, from $d=3$ \cite{CAdS2,CdSgeo} to any dimensions,  is somewhat natural since we used Cartesian coordinates. Nevertheless, our method of exploiting conserved quantities  was more productive in studying the Boltzmann equation allowing us to derive the general form of the distributions of the Boltzmann-Marle model on the dS and AdS spacetimes of any dimensions. This represent the principal new results that may be used in further applications to concrete physical systems of various dimensions.

\appendix

\section{Static charts}

The Cartesian coordinates of the standard static charts are related to our coordinates used here as
\begin{equation}
x_{s\pm}^i=\frac{x^i}{\sqrt{1\pm\omega^2 x^2}}\,,
\end{equation} 
such that the local temperatures (\ref{temp}) take the form
\begin{equation}
\beta_E^{\pm}=\beta_0 \sqrt{1\mp  \omega^2 x_{s\pm}^2}\,, 
\end{equation}
we found in Ref. \cite{AC}.
The line elements in the standard static charts with Cartesian coordinates of $M_+$ and $M_-$ read
\begin{equation}
ds^2_{\pm}=(1\mp x_{s\pm}^2)dt_{\pm}^2-\left[\delta_{ij}\pm\omega^2\frac{x^i_{s\pm}x^j_{s\pm}}{1\mp x_{s\pm}^2}\, dx^i_{s\pm}dx^j_{s\pm} \right]
\end{equation}
giving the familiar expressions in spherical coordinates.

\section*{Acknowledgments}
This work is partially supported by a grant of  the Romanian Ministry of Research and Innovation, CCCDI-UEFISCDI, project number  PN-III-P1-1.2-PCCDI-2017-0371.


\begin{thebibliography}{20}

\bibitem{cac1} 
S. Cacciatori, V. Gorini, A. Kamenshchik, {\em Annalen der Physik} {\bf 17} (2008) 728.

\bibitem{cac2} 
S. Cacciatori, V. Gorini, A. Kamenshchik and U. Moschella U,  {\it Class. Quantum Grav.} {\bf 25}  (2008) 075008. 

\bibitem{CdS1}
I. I. Cot\u aescu, {\em Eur. Phys. J. C} {\bf 77} (2017)  485; arXiv:1701.08499. 

\bibitem{CdS2}
I. I. Cot\u aescu, {\em Eur. Phys. J. C} {\bf 78} (2018)  95; arXiv:1708.06638. 

\bibitem{CdSgeo}
I. I. Cot\u aescu, {\em Mod. Phys. Lett. A} {\bf 32} (2017) 1750223,
arXiv:1711.02956. 

\bibitem{CAdS1}
I. I. Cot\u aescu,  {\em Phys. Rev. D} {\bf 95} (2017) 104051.

\bibitem{CAdS2}
I. I. Cot\u aescu,  {\em Phys. Rev. D} {\bf 96} (2017) 044046.

\bibitem{P1}
E. van Beveren, G. Rupp, T. A. Rijken and C. Dullemond, {\it Phys. Rev. D} 
{\bf  27}  (1983) 1527.

\bibitem{P2}
C. Dullemond and E. van Beveren, {\it Phys. Rev. D} {\bf 28}  (1983) 1028.

\bibitem{Cer}
C. Cercignani and G. M. Kremer, {\em The relativistic Boltzmann
equation: theory and applications},( Birkh\" auser Verlag,
Basel, Switzerland 2002).

\bibitem{G1}
S. S. Gubser, {\em Phys. Rev. D} {\bf 82}  (2010) 085027.

\bibitem{G2}
S. S. Gubser and A. Yarom, {\em Nucl. Phys.} {\bf B846} (2011) 469.

\bibitem{B1}
G. Denicol, U. Heinz, M. Martinez, J. Noronha and M. Strickland,  {\it Phys. Rev. D} {\bf 90}  (2014) 125026.

\bibitem{B2}
D. Bazow, G. S. Denicol, U. Heinz, M. Martinez, and J. Noronha, {it Phys. Rev. Lett}, {\bf  116}  (2016) 022301.

\bibitem{AC}
V. E. Ambrus and I.  I. Cot\u aescu, {\em Phys. Rev. D} {\bf 94} (2016) 085022.

\bibitem{ES}
I. I. Cot\u aescu, {\em J. Phys. A: Math. Gen.} {\bf 33}  (2000) 9177.

\bibitem{Nach}
O. Nachtmann, {\em Commun. Math. Phys.} {\bf 6} (1967) 1.






 
 
\end{thebibliography}
\end{document}